\documentclass[pra,a4paper,final,twocolumn,showpacs]{revtex4}
\def\c60{C$_{60}$}
\def\na#1{Na$_{#1}$}
\usepackage{amssymb}
\usepackage{amsmath}
\begin{document}
\title{Wetting to Non-wetting Transition in Sodium-Coated C$_{60}$}
\author{J. Roques$^{(1,2)}$, F. Calvo$^{(1)}$, F. Spiegelman$^{(1)}$, and
C. Mijoule$^{(2)}$}
\affiliation{(1) Laboratoire de Physique Quantique, IRSAMC, Universit\'e Paul
Sabatier, 118 Route de Narbonne, F31062 Toulouse Cedex \\
(2) CIRIMAT, Universit\'e Paul Sabatier and Institut National Polytechnique,
118 Route de Narbonne, F31062 Toulouse
Cedex}
\begin{abstract}
Based on {\em ab initio}\/ and density-functional theory calculations, an
empirical potential is proposed to model the interaction between a fullerene
molecule and many sodium atoms. This model predicts homogeneous coverage of
\c60 below 8 Na atoms, and a progressive droplet formation above this size.
The effects of ionization, temperature, and external electric field indicate
that the various, and apparently contradictory, experimental results can indeed
be put into agreement.
\end{abstract}
\pacs{61.48.+c,36.40.Qv,34.70.+e}
\maketitle

Fullerene molecules are now commonly used as building blocks of complex
materials having unusual physical and chemical properties \cite{c60review}.
The possibilities offered by molecules such as \c60 or C$_{70}$ in terms of
electric or optical devices have laid ground for the rapidly expanding
nanotechnology area. The interaction between fullerenes and alkali atoms
has attracted a lot of attention after the discovery of
superconductivity in the K$_3$\c60 solid \cite{hebard}. Sodium-\c60
compounds have also received much attention \cite{na4c60,na2c60}.
More recently,
several groups investigated experimentally the gas phase properties of
exohedral alkali-\c60 compounds \cite{wang,martin,palpant}, to further
predict the solid state of alkali-doped fullerite.

Using mass spectrometry, Martin and coworkers \cite{martin} inferred from
the most stable ``magic numbers'' that sodium covers the \c60 molecule in
a continuous and homogeneous fashion, and that metallic bonding starts above
6 metal atoms. Palpant and coworkers \cite{palpant} deduced from photoelectron
spectroscopy measurements that coating proceeds by trimers units rather than
single atoms. They also estimated that metallic bonding only appears at
$n=13$ sodium atoms.
Very recently, Dugourd and coworkers \cite{dugourd} measured the electric
polarizability of \na{n}\c60 clusters in the range $1\leq n\leq 34$. They
concluded that a sodium droplet is formed on the surface of the fullerene.
These apparent contradictory results are surprising, as far as they rely on
similar experimental conditions but different interpretations.

The situation is in fact even more intricate due to the variety of
theoretical conclusions on the very same systems
\cite{rubio,hira1,hira2,hamamoto}. With the exception
of the study by Rubio and coworkers \cite{rubio}, which assumes complete
wetting of \c60 by sodium in a continuous two-shell jellium description, the
electronic structure calculations by Hira and Ray \cite{hira1,hira2}, and by
Hamamoto {\em et al.} \cite{hamamoto} reach different conclusions as to whether
the \na{2} molecule remains as a dimer or dissociates into atoms, which locate
on opposite sides of
\c60. At the unrestricted Hartree-Fock level, Hira and Ray find that charge
transfer is negligible in Na\c60, in apparent contradiction with experiments
\cite{palpant,dugourd}, and also that \na{2} remains as a dimer loosely bound
to \c60. On the other hand, the more realistic density-functional theory (DFT)
calculations performed by Hamamoto and coworkers \cite{hamamoto} at the local
density approximation level tend to favor regular coating, preferentially by
trimers, in agreement with the picture of Palpant {\em et al.} \cite{palpant}.
However, all these {\em ab initio} investigations could only focus on very
limited sets of chosen geometries, simply because global optimization on such
complex energy landscapes is far beyond the current possibilities of fully
quantum mechanical approaches. In addition, statistical and temperature effects
still remain a major limitation in large scale simulations based on first
principles.

To overcome the above difficulties, we have constructed an empirical, {\em ab
initio}\/ based atomistic model allowing systematic static and dynamical
investigations in a large size range. Briefly, given the geometry ${\bf R}=\{
{\bf r}_i, {\bf r}'_j\}$ of the \na{n}\c60$^{Q}$ system, with ${\bf r}_i
\in$ \na{n} and ${\bf r}'_j\in$ \c60, the total potential energy is written as
\begin{equation}
V({\bf R}) = V_{\rm Na_{\it n}}({\bf r}_i) +
V_{\rm C_{60}}({\bf r}'_j) + V_{\rm
inter}({\bf R}) + V_{\rm Coul}({\bf R}),
\label{eq:vtot}
\end{equation}
where $V_{\rm Na_{\it n}}$, $V_{\rm C_{60}}$, $V_{\rm inter}$, and $V_{\rm
Coul}$ are the
energies of sodium, \c60, covalent interactions between Na and \c60, and
Coulomb interactions, respectively. Metallic forces in sodium are described by
a simple many-body potential in the second-moment approximation of the
electronic density of states in the tight-binding theory \cite{sma}. Covalent
forces in \c60 were most often neglected, as the temperatures of interest are
low enough ($<300$~K) to keep the fullerene rigid. We occasionaly checked the
rigidity of \c60 in some optimal structures with the Tersoff potential
\cite{tersoff}. The covalent interaction between sodium and carbon atoms was
modelled as a pairwise repulsion term taken from {\em ab initio}\/ calculations
of the {}$^4\Pi$ state of the NaC molecule, and expressed
as $V_{\rm NaC}^{\rm covalent}(r)=D\exp
(-\alpha r)$. To account for charge transfer, and to quantify the extent of
ionic interaction in the system, fluctuating charges $Q=\{ q_i, q'_j\}$, with
$q_i\in$ \na{n} and $q'_j\in$ \c60, were added to all atoms in the system. For
any geometry ${\bf R}$, the charges are determined in order to minimize the
global Coulomb interaction $V_{\rm Coul}$, which is equivalent to equalizing
the effective electronegativities in the sense of Sanderson \cite{sanderson}:
\begin{eqnarray}
V_Q({\bf R}) &=& \sum_i \varepsilon_{\rm Na}q_i + \frac{1}{2} H_{\rm Na}q_i^2
+ \sum_j \varepsilon_{\rm C} q'_j + \frac{1}{2}H_{\rm C}{q'_j}^2 \nonumber \\
&& + \sum_{i<i'} J_{ii'} q_i q_{i'} +\sum_{j<j'}J_{jj'} q'_j q'_{j'}\nonumber\\
&& + \sum_{i,j} J_{ij}q_i q'_j + \lambda\left( \sum_i q_i + \sum_j q'_j -Q
\right). \label{eq:vq}
\end{eqnarray}
In this equation, $i$ and $j$ label sodium and carbon atoms, respectively.
$\varepsilon_{\rm Na}$ and $\varepsilon_{\rm C}$ are the respective
electronegativities of sodium and carbon, $H_{\rm Na}$ and $H_{\rm C}$ their
`hardnesses', $J_{ij}$ the Coulomb interactions. The hardnesses parameters
correspond to the on-site repulsion in the Hubbard model.
The latter quantities were
extracted from {\em ab initio}\/ calculations of the Coulomb integrals. Due to
the finite delocalization of the electrons, $J_{ij}$ does not diverge at short
distances. We have used the simple expression $J_{ij}(r) = \left[ r^2 +
H_{ij}^{-2}\exp(-\gamma_{ij} r^2)\right]^{-1/2}$ to model the Coulomb parts of
the Na--Na, C--C, and Na--C interactions. The hardnesses parameters $H_{\rm
Na}$ and $H_{\rm C}$ are the $r\to 0$ limits of $J_{\rm Na-Na}$ and $J_{\rm
C-C}$. Finally, the Lagrange multiplier $\lambda$ in (\ref{eq:vq}) ensures that
the system carries a global charge $Q$. This electrostatic model was further
improved by including 90 fixed charges, 60 on each carbon site with value
$\delta q$, 30 on the middle of each C=C bond with value $-2\delta q$. This
was previously used by Schelkacheva Tareyeva in a study of bulk \c60
\cite{fixedq}.

The dynamical use of the full potential energy function is greatly facilitated
by considering an extended Lagrangian where the fluctuating charges are treated
as independent variables \cite{rick}, in a way similar to the Car-Parrinello
scheme \cite{carpar}.
This empirical model has 15 parameters, namely the 5 parameters of the
SMA potential, the 2 parameters of the covalent Na--C interaction, the
electronegativity difference $\varepsilon_{\rm C}-\varepsilon_{\rm Na}$, the 6
parameters of the Coulomb interactions (including hardnesses), and the fixed
charge $\delta q$. These parameters were obtained by minimizing an error
function $\chi^2$ to reproduce several properties independently evaluated by
{\em ab initio}\/ and DFT calculations \cite{roques} with the B3LYP nonlocal
functional \cite{b3lyp}. These properties are the charge transferred from
sodium (0.87e) and the electric dipole (14.5~D) in Na\c60, the energy
difference
between the two \na{2}\c60 isomers with sodium atoms on adjacent or opposite
hexagonal sites ($\Delta E=-0.35$~eV), and the binding energy and equilibrium
distance in \na{2} (resp. $D_e=$0.73~eV and $R_e=$3.08~\AA).
It should be noticed that
these theoretical values agree well with the experimental data available
\cite{dugourd}. We also added to the error function a penalty term to reproduce
the experimental electric polarizability of the \c60 molecule (76.5~\AA$^3$
\cite{polarc60}).

\begin{figure}[htb]
\vbox to 6.4cm{
\includegraphics{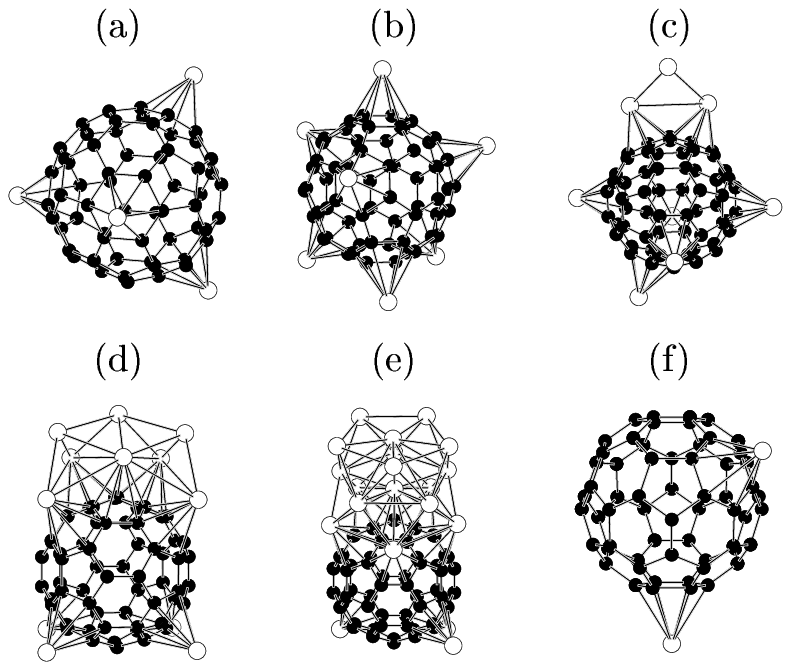}
\vfill}
\caption{Some optimal structures of \na{n}\c60 clusters. (a) $n=4$; (b) $n=7$;
(c) $n=8$; (d) $n=12$; (e) $n=20$. The lowest energy structure for \na{2}\c60
within an electric field $E=2.55\times 10^8$ V.m$^{-1}$ if represented in (f).}
\label{fig:struct}
\end{figure}

The lowest energy structures of \na{n}\c60 clusters were determined using a
variant of the basin-hopping, or Monte Carlo + minimization algorithm
\cite{bh}, in which each sodium atom is offered a probability to rotate freely
over the \c60 surface, in addition to the usual random displacements moves.
Some of the optimal structures are represented in Fig.~\ref{fig:struct} for
the sizes $n=4$, 7, 8, 12, and 20. Up to seven sodium atoms, the empirical
model predicts ``wetted'' structures, where each alkali atom stays over an
hexagonal site. In this regime, the Coulomb repulsion between the positive
charges carried by the sodium atoms is minimized by keeping the atoms as far
away as possible from each other. At the stage of 7 atoms, there is only little
space left on the \c60 surface to add an extra eighth isolated atom,
and the energy
gained in creating Na--Na bonds gets larger than the electrostatic penalty.
This embryonic droplet further grows as new sodium atoms are incorporated, and
progressively spreads over the \c60 surface. Growth occasionally proceeds by
removing one of the remaining isolated atoms on the side opposite to the
droplet, and may eventually end in a single, big droplet. The largest system
considered here, \na{30}\c60, however, still shows one main droplet and two
isolated atoms.

\begin{figure}[htb]
\vbox to 6.2cm{
\includegraphics{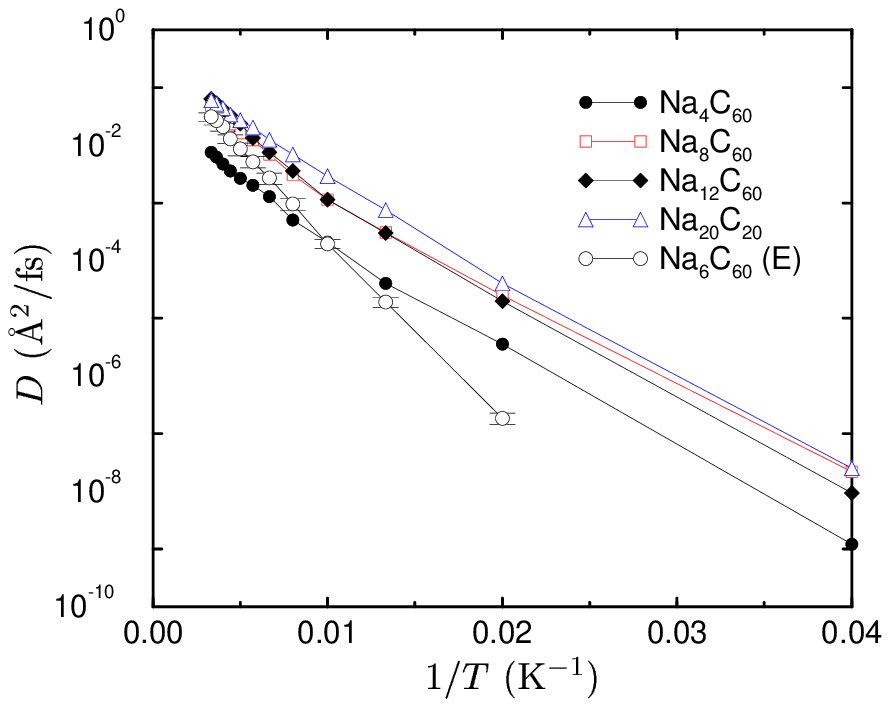}
\vfill}
\caption{Arrhenius plots of the diffusion constants of some \na{n}\c60
clusters, for $n=4$, 8, 12, 20 (no field), and 6 within an electric field.}
\label{fig:D}
\end{figure}

\begin{figure}[htb]
\vbox to 6.2cm{
\includegraphics{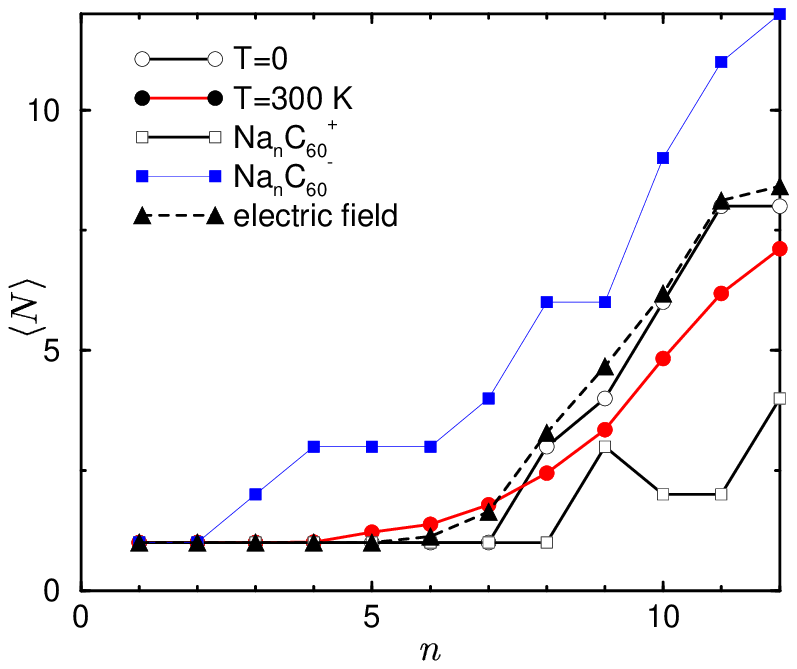}
\vfill}
\caption{Size of the largest fragment of sodium atoms in \na{n}\c60 clusters
in various conditions, and for neutral or charged molecules.}
\label{fig:frag}
\end{figure}

To provide a more complete interpretation of the experiments, we have
considered the separate effects of temperature, ionization, and a possible
external electric field on the transition between coated and segregated
morphologies. Equilibrium dynamics at constant temperature was achieved by
supplementing the extended Lagrangian with two sets of Nos\'e-Hoover chain
thermostats \cite{nose}, one set at the desired ionic temperature $T$, one
set to keep the average kinetic energy of the charges to a low value
$T^*=T/100$. Thermostating the charges also prevents the fictitious dynamics
to diverge too much from the adiabatic, Born-Oppenheimer type dynamics
\cite{sprik}.
Diffusion constants were estimated from the slopes of the average mean square
atomic displacements versus time in a set of 500 simulations at various
temperatures in the range 25~K$\leq T\leq$300~K, with increments 25~K. 
In Fig.~\ref{fig:D}, these diffusion constants are represented for the sizes
4, 8, 12, and 20 as a function of $1/T$ in an Arrhenius plot. These plots
provide us with estimates of the corresponding activation barriers, for
which we find $A\simeq 400$~K for all four sizes. At room temperature,
sodium atoms thus show a significant mobility over the \c60 surface. This
agrees with the recent experimental findings on K\c60 \cite{kc60}. To quantify
the effect of temperature on the morphology in \na{n}\c60, we have calculated
the average size of the largest sodium fragment. A fragment is here
defined as a set of connected atoms, {\em i.e.} having at least one neighbor
at a distance shorter than 8 bohr. The same geometrical observable was also
computed at $T=0$ for neutral or singly charged molecules. In the case of
charged systems, global optimization of \na{n}\c60$^+$ and \na{n}\c60$^-$
without electric field was performed for sizes near the crossover, $n\leq 12$.
Lastly, we have compared the results without electric field and those with a
homogeneous field of magnitude $E=2.55\times 10^8$~V.m$^{-1}$ relevant to the
experiments of Dugourd {\em et al.} \cite{dugourd}, and at zero temperature.
The corresponding data were averaged on 30 different orientations of the field,
relative to the \c60 geometry. Global optimization was carried out for each of
these orientations. The variations of the
largest fragment size $\langle N\rangle$ against the total number of sodium
atoms $n$ are plotted in Fig.~\ref{fig:frag}.
The rise of $\langle N\rangle$ at the crossover size $n^*$ marks the onset of
droplet formation. As temperature increases, isolated atoms are more likely to
meet either with each other or with the preexisting droplet.
This lowers $n^*$ by a visible amount. However, 300~K is a relatively high
temperature for sodium clusters lying usually above their melting points
\cite{haberland}. Thus, external bonds are often broken in the course of the
dynamics,
which results in lowering $\langle N\rangle$ at larger $n$. Ionizing the
cluster can also strongly affect the crossover size $n^*$: negatively
charged molecules are more stable in segregated form due to a much lower
Coulomb repulsion of the alkali atoms, and positively charged molecules are
less stable and remain homogeneously coated up to $n=10$.

At the experimental value, a finite electric field disfavors regular location
of the charges over the nearly spherical \c60 surface. This yields more
frequent metallic bonds. In addition, the lowest energy structures found for
all field directions can also change significantly. For instance, as seen in
Fig.~\ref{fig:struct}, two sodium atoms get closer to one another, yet they
still do not form a bond at 0~K.

\begin{figure}[htb]
\vbox to 6.2cm{
\includegraphics{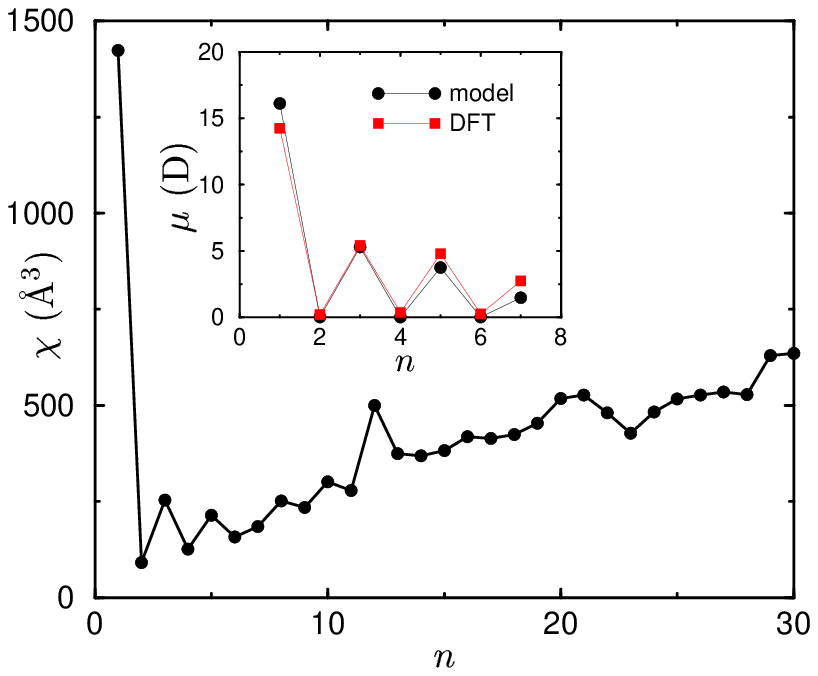}
\vfill}
\caption{Electric susceptibilities at 300~K. Inset: electric dipole moment
and comparison with DFT calculations.}
\label{fig:polar}
\end{figure}

Electric susceptibilities $\chi$ have been estimated with the present model
using the high-temperature approximation to the Langevin theory for rigid
dipoles undergoing orientational thermal fluctuations:
$\chi = \alpha + \mu^2/3k_{\rm B}T$. In this formula
$\alpha$ is the static polarizability, $k_{\rm B}$ is the Boltzmann constant,
and $\mu$ the electric dipole. The variations of $\chi$ with cluster size $n$
are shown in Fig.~\ref{fig:polar}.
They qualitatively resemble the experimental data of Dugourd {\em et al.}
\cite{dugourd}, but lie somewhat lower than the actual measured values. This
may first come from the parameterization. We tested this hypothesis by
comparing the predictions of the present empirical model to single-point DFT
calculations performed at the same geometry. The electric dipoles given in the
inset of Fig.~\ref{fig:polar} show good agreement between the empirical and
electronic structure estimates. We also note that the strong variations in the
dipole reflect the changes in the molecular symmetry. It is also worth pointing
out that the model predicts an energetical ordering of the isomers essentially
similar to that found by Hamamoto {\em et al.} \cite{hamamoto} for the very
restricted set of structures considered by these authors. Therefore, the
quantitative difference between our results and the experimental data more
likely comes from the Langevin formula, strictly valid for rigid dipoles. This
approximation may be insufficient to describe the dipole fluctuations
associated with the floppy character of the metal droplet. However, regarding
the difference between the present results and the data obtained in Ref.
\onlinecite{dugourd}, adding several isolated atoms on the opposite
side of the growing droplet decreases the dipole moment, hence the
susceptibility. A fully dynamical simulation would
probably be required to get a more quantitative agreement.

The present investigation shows that the appearance of metallic bonding
corresponds to seeding a droplet, which progressively grows and captures
remaining isolated atoms. This picture partly reconciles the apparently
contradictory experimental interpretations \cite{martin,palpant,dugourd}.
The empirical potential was fitted to reproduce electronic structure
calculations as well as some experimental data.
By allowing large scale simulations
and structural sampling unaccessible to quantum mechanical studies, we could
estimate the crossover size between wetted and nonwetted morphologies to
be located near 8 atoms. This is also the size range where the metallic
transition can be estimated to occur.
We have also been able to investigate how the crossover
size depends on various effects such as charge or temperature.

The current limitations of the present empirical model are mostly due to the
lack of an explicit account of electronic structure. Even though
Coulomb repulsion should play an important role when only few sodium atoms
are effectively charged but close to each other, shell closing can enhance
stability in a local way. In particular, this could help in explaining why
charged trimers remain stable. Such quantum effects are obviously beyond the
fluctuating charges model, especially with its current training on purely
neutral molecules. A next step could be a simple quantum tight-binding
approach, which would include both covalent and charge transfer effects.
However, a realistic tight-binding Hamiltonian for the \na{n}\c60 system would
require $240+n$ electronic states (4 $s+p$ electrons for each carbon atom, 1
$s$ valence for each sodium atom). This is still too heavy for exhaustive
sampling and large scale simulations. In addition, precise quantum effects are
essentially expected at small sizes, and the role of a single charge on the
structure and stability should decrease as the sodium droplet grows.

For large sizes, 
it would be interesting to study in more details how the physical and chemical
properties of the sodium droplet are modified by the interaction with the \c60
molecule. The influence of the \c60 deformation and vibration on the metal
cluster dynamics could also be an important issue.
Use of the present model to treat other metals could help in
understanding the differences observed in the experiments
\cite{palpant,dugourd,antoine}. Further extensions, including the treatment of
endohedral or exohedral fullerenes, nanotubes or surfaces with materials
possibly other than metals \cite{c60f48,stuart}, could also be carried out.

The density functional calculations were performed using the GAUSSIAN98
software package \cite{gaussian}.
We thank CALMIP for a generous allocation of computer resources.

\end{document}